\documentclass[twocolumn,conference]{IEEEtran}

\usepackage{color}
\usepackage{multirow,amssymb} 
\usepackage{algorithmic}
\usepackage{cite}

{\bfseries}{\itshape}
\newtheorem{proposition}{\bfseries\textit{Proposition}}
\newtheorem{example}{\bfseries\textit{Example}}

\begin{document}

\title{Connecting Randomized Response, 
Post-Randomization, Differential Privacy and $t$-Closeness 
via Deniability and Permutation} 

\author{
\IEEEauthorblockN{Josep Domingo-Ferrer~\IEEEmembership{Fellow,~IEEE} and Jordi Soria-Comas}
\IEEEauthorblockA{Universitat Rovira i Virgili\\
Dept. of Computer Science and Mathematics\\
UNESCO Chair in Data Privacy\\
Av. Pa\"{\i}sos Catalans 26, 43007 Tarragona, Catalonia\\
Email: \{josep.domingo,jordi.soria\}@urv.cat}}
	\maketitle

	\begin{abstract}
We explore some novel connections between the main privacy models in use
and we recall a few known ones. We show these models to be more related
than commonly understood, around two main principles: deniability
and permutation. In particular, randomized response 
turns out to be very modern in spite of it having been introduced 
over 50 years ago: it is a local anonymization method 
and it allows understanding the protection offered
by $\epsilon$-differential privacy when $\epsilon$ is increased
to improve utility. A similar understanding on the effect of 
large $\epsilon$ in terms of deniability 
is obtained from the connection between
$\epsilon$-differential privacy and $t$-closeness.
Finally, 
the post-randomization method (PRAM) is shown to be viewable
as permutation and to be  
connected with randomized response and differential privacy.
Since the latter is also connected with $t$-closeness, 
it follows that the permutation 
principle can explain the guarantees offered by all those models. 
Thus, calibrating permutation is very relevant in anonymization, 
and we conclude by sketching two ways of doing it.
	\end{abstract}
	
	\section{Introduction}

Privacy models in the literature can be split into four main families,
by order of appearance: i) randomized response (1965); 
ii) $k$-anonymity (1998) and its 
extensions ($l$-diversity (2006), $t$-closeness (2007), etc.); 
iii) differential privacy (2006); 
and iv) the permutation paradigm (2016).

Although the above models were proposed over a 50-year span 
and were initially believed to be fundamentally different from each 
other, some connections between pairs of them have been found
in the literature. Such connections can be useful to the anonymization
practitioner because they give more insight into the risk-utility trade-offs
incurred when choosing a certain parameterization for a specific model.

We explore in this paper several connections 
from a new perspective.
Section~\ref{rrpram} connects
randomized response and post-randomization (PRAM) in terms of plausible 
deniability; thus, randomized response is not
only an anonymization method {\em avant la lettre}, 
but more interestingly a {\em local} version of 
the PRAM anonymization method. 
Section~\ref{rrdp} dwells on the known link
between randomized response and differential privacy,
by taking a new approach based on information theory 
and deniability; in particular, our analysis allows
understanding the effect on privacy 
of increasing the $\epsilon$ parameter of differential privacy
(in an attempt to improve utility).
Section~\ref{dpt} recalls the connection between 
differential privacy and $t$-closeness that we demonstrated
in a previous paper; the novelty is that our deniability-based
analysis permits using $t$-closeness to assess whether
a large $\epsilon$ in differential privacy provides enough protection.
Section~\ref{pramperm} views PRAM in terms of the permutation 
paradigm. The lessons learned from the previous sections 
pave the way to calibrating anonymization in terms
of permutation, whatever the 
privacy model in use; this is dealt with in Section~\ref{calibrate}.
Finally, Section~\ref{conclude} contains conclusions
and future research issues. 

\section{Randomized response, plausible deniability and PRAM}
\label{rrpram}

Randomized response (RR,~\cite{citerr1965,citerr1969}) 
is a mechanism that respondents 
to a survey can use to protect their privacy 
when asked about the value of sensitive attribute
({\em e.g.} did you take drugs last month?).
The interesting point is that the data collector can still 
estimate from the randomized responses 
the proportion of each of the possible {\em true} answers
of the respondents.

Let us denote by $X$ the attribute containing the answer
to the sensitive question. If $X$ can take $r$ possible
values, then 
the randomized response $Y$ reported 
by the respondent instead of $X$ follows a $r \times r$ 
matrix of probabilities
\begin{equation}
\label{probmatrix}
{\bf P} = \left( \begin{array}{ccc} p_{11} & \cdots & p_{1r} \\
\vdots & \vdots & \vdots \\
p_{r1} & \cdots & p_{rr} \end{array}\right) 
\end{equation}
where $p_{uv} = \Pr(Y = v| X=u)$, for $u,v \in \{1,\ldots,r\}$
denotes the probability that the randomized response
is $v$ when the respondent's true attribute value is $u$.

Let $\pi_1,\ldots,\pi_r$ be the proportions of respondents whose true values 
fall in each of the $r$ categories of $X$ and let 
$\lambda_v = \sum_{u=1}^r p_{uv} \pi_u$ for $v=1,\ldots, r$, 
be the probability of the reported value $Y$ being $v$.
If we define ${\bf \lambda} = (\lambda_1,\ldots,\lambda_r)^T$ 
and ${\bf \pi} = (\pi_1,\ldots, \pi_r)^T$, it holds that
${\bf \lambda} = {\bf P}^T {\bf \pi}$. Furthermore, 
if $\hat{\bf \lambda}$ is the vector of sample proportions
corresponding to ${\bf \lambda}$ and ${\bf P}$ is nonsingular,
in Chapter 3.3 
of~\cite{Chaud88} it is proven that an unbiased estimator
${\bf \pi}$ can be computed as
\begin{equation}
\label{unbiased}
\hat{\bf \pi}  = ({\bf P}^T)^{-1} \hat{\bf \lambda}
\end{equation}
and they also provide an unbiased estimator of the dispersion matrix.

\subsection{The privacy model of randomized response: plausible deniability}
\label{deny}

Even if the concept of privacy model (that is,
of {\em ex ante} privacy guarantee) was introduced
by $k$-anonymity~\cite{samarati98} three decades 
after RR had appeared, RR has an implicit privacy 
model. The privacy guarantees RR offers to respondents
are {\em plausible deniability} and {\em secrecy}, as we next analyze. 

For each possible value $v$ of the reported
attribute $Y$, by the Bayes' formula we have 
\begin{eqnarray}
\hat{p}_{vu}&=&\Pr(X=u|Y=v)\nonumber\\
&=&\frac{\Pr(Y=v|X=u) \Pr(X=u)}{\sum_{u'=1}^r\Pr(Y=v|X=u')\Pr(X=u')}\nonumber\\
&=&\frac{p_{uv} \pi_u}{\sum_{u'=1} p_{u'v} \pi_{u'}}\label{eqbayes}.
\end{eqnarray}

\paragraph{Deniability} Imagine that $u$ is an embarrassing value 
of $X$ (like a value 
``Yes'' for an attribute $X$ denoting whether
drugs were taken last month). 
As long as $\Pr(X=u|Y=v) < 1$, the respondent can deny
to have $X=u$. 

Actually, the more similar the probabilities 
$\hat{p}_{vu}$ corresponding  
to a given answer $Y=v$, the higher the deniability.
Therefore, given a reported
value $Y=v$, we can measure deniability as a 
conditional Shannon entropy 
\begin{equation}
\label{deniability}
H(X|Y=v) = - \sum_{u=1}^r \hat{p}_{vu} \log_2 \hat{p}_{vu},
\end{equation} 
whose maximum value is $\log_2 r$, corresponding 
to the case when $\hat{p}_{uv}$ takes the same value $1/r$
for all $u=1,\ldots, r$. 

\paragraph{Perfect secrecy}
In the special case in which the probabilities
within each column of ${\bf P}$ are identical 
(although perhaps different columns contain different
probabilities), Expression (\ref{eqbayes}) tells us 
that $\hat{p}_{vu}= \pi_u$, for $u,v \in \{1,\ldots, r\}$.
Therefore, using Expression (\ref{deniability}) we get 
that $H(X|Y=v)=H(X)$ for any $v$, and thus  $H(X|Y) = H(X)$,
which implies {\em perfect secrecy} in the Shannon sense~\cite{Shan49}: 
the reported answer $Y$ gives no information at all on the 
real value of $X$.

Unfortunately, the price paid for perfect secrecy
is high in terms of utility: when
the probabilities within two or more columns of ${\bf P}$ are
identical, matrix ${\bf P}$ is singular,
and therefore the unbiased estimator of Expression (\ref{unbiased})
cannot be computed.

Finally, having $H(X|Y=v)=H(X)$ for some $v$ yields an uninformative 
reported value. 
It is possible to go beyond that and make $v$ misinformative, by ensuring
that  
$H(X|Y=v)>H(X)$. However, since $H(X|Y)\le H(X)$, 
if we increase $H(X|Y=v)$ to make it greater than $H(X)$
for some $v$, we are forced to decrease $H(X|Y=v')$ for other
values $v'\neq v$. 

\subsection{Randomized response: a local version of PRAM}

The matrix in Expression (\ref{probmatrix}) looks 
exactly as the transition matrix used in the post-randomization
method (PRAM) proposed by~\cite{PRAM}. As pointed out 
in~\cite{tesiardo}, the main difference between RR
 and PRAM is who performs the randomization: 
whereas in RR it is the respondent before 
delivering her response, in PRAM it is the data
controller after collecting all responses (hence the name
post-randomization). Therefore, RR is a local version of PRAM 
anonymization, which is not without merit:
 when RR was invented, the notion of anonymization did not exist,
let alone the notion of local anonymization.

\section{Randomized response and differential privacy}
\label{rrdp}

A randomized query function $\kappa$ gives $\epsilon$-differential
privacy~\cite{Dwork2006} if, for all
data sets $D_{1}$, $D_{2}$ such that one can
be obtained from the other by modifying a single record,
and all $S\subset Range(\kappa)$, it holds 
\begin{equation}
\label{dp1}
\Pr(\kappa(D_{1})\in S)\le\exp(\epsilon)\times \Pr(\kappa(D_{2})\in S.
\end{equation}
In plain words, the presence or absence of any single record
is not noticeable (up to $\exp(\epsilon)$) when seeing the outcome of the query.
Hence, this outcome can be disclosed without
impairing the privacy of any of the potential respondents
whose records might be in the data set.
A usual mechanism 
to satisfy Inequality (\ref{dp1}) is to add noise to the 
true outcome of the query, in order to obtain an outcome of $\kappa$ 
that is a noise-added version of the true outcome. The 
smaller $\epsilon$, the more noise is needed to 
make queries on $D_1$ and $D_2$ indistinguishable up 
to $\exp(\epsilon)$.

In~\cite{Wang16,Wang14}, a connection between RR
and differential privacy is established: 
RR is $\epsilon$-differentially private if 
\begin{equation}
\label{eqdp}
 e^\epsilon \geq \max_{v=1,\ldots,r} \frac{\max_{u=1,\ldots,r} p_{uv}}{\min_{u=1,\ldots,r} p_{uv}}.
\end{equation}
The rationale is that the values in each column $v$ ($v \in \{1,\ldots,r\}$)
of matrix ${\bf P}$ correspond to the probabilities of the reported
value being $Y=v$, given that the real value is $X=u$ 
for $u \in \{1,\ldots,r\}$. 
Differential privacy requires that the 
maximum ratio between the probabilities in a column be bounded 
by $e^{\epsilon}$, so that the 
influence of the real value $X$ on the reported value $Y$ 
is limited. Thus, the reported value can be released
with limited disclosure of the real value.

\subsection{Connection with entropy}

The smaller the value of $\varepsilon$ in Inequality (\ref{eqdp}), 
the more similar the probabilities $p_{uv}$ ($u=1,\ldots,r$) 
in the columns of 
${\bf P}$, for columns $v=1$ to $r$. 
In turn, the more similar these probabilities, 
the more $\hat{p}_{vu}$ approaches $\pi_u$, as it 
can be seen from Expression (\ref{eqbayes}). 

In the extreme case,
when $\epsilon=0$, Inequality (\ref{eqdp}) forces 
the probabilities within each column to be identical to each other
(although not necessarily equal to probabilities in other columns).
As shown in Section~\ref{deny}, this implies $H(X|Y)=H(X)$.
Therefore, {\em if randomized response 
achieves the strictest differential privacy ($\epsilon=0$),
it also achieves 
perfect secrecy in the Shannon sense.}

Yet, one might argue
that conditional entropy $H(X|Y)$ 
captures more information than Inequality (\ref{eqdp}), 
because it takes
all probabilities into account, whereas 
 Inequality (\ref{eqdp}) (and hence the above 
connection of RR with differential privacy) 
only takes into account the maximum and the minimum probabilities
in each column.  

\subsection{Explaining large $\epsilon$ in terms of deniability}
\label{largepsilon}

One of the shortcomings of the differential privacy model is that
its {\em ex ante} privacy guarantee is only intuitive
when $\epsilon$ is very small.
Specifically, when one
takes not-so-small $\epsilon$ values 
in Expression (\ref{dp1}) to preserve more utility, the 
privacy guarantee is hard to explain: is $\epsilon$ is 
not that small, one cannot guarantee 
any more that the presence
or absence of any single record/respondent 
is really unnoticeable.

The connection with randomized response and hence deniability
given by Inequality (\ref{eqdp}) is useful to gain an intuition
on what large $\epsilon$ implies. We illustrate
this in the following example.

\begin{example}
\label{ex1}
If one takes $\epsilon=2$, this 
means that in some columns of ${\bf P}$ the ratio 
between the largest probability and the smallest probability
may be as large as $e^2=7.389$. In particular, if say, 
$r=2$, one might have a column $v$ with one probability
$p_{1v}=0.7389$ and the other probability $p_{2v}=0.1$. 
In this situation, if the prior probabilities of the two values of $X$
are similar,
 Expression (\ref{eqbayes}) tells us that, after reporting $Y=v$, 
the most likely value of $X$ 
 is 1 and there is
only a narrow margin for denying it. {\em This clearly shows 
that $\epsilon=2$ does not seem to provide enough privacy.}
\end{example}

\section{Differential privacy and $t$-closeness}
\label{dpt}

Given two random distributions
$F_{1}$ and $F_{2}$ taking values in a discrete set $\{x_{1},x_{2},$
$\cdots,$ $x_t\}$, consider the following distance between them: 
\begin{equation}
d(F_{1},F_{2})=\max_{i=1,2,\cdots,t}\left\{ \frac{\Pr_{F_{1}}(x_{i})}{\Pr_{F_{2}}(x_{i})},\frac{\Pr_{F_{2}}(x_{i})}{\Pr_{F_{1}}(x_{i})}\right\} .\label{eq:distance}
\end{equation}
In Expression (\ref{eq:distance}),
we take the quotients of probabilities to be zero if both 
$\Pr_{F_{1}}(x_{i})$ and $\Pr_{F_{2}}(x_{i})$ are zero, 
and to be infinity if only the denominator is zero.

In~\cite{knosys}, it is proven that, if an anonymized static data
set satisfies $\exp(\varepsilon/2)$-closeness~\cite{litclose} (an extension
of $k$-anonymity) when the distance between
the distribution of the sensitive attributes over entire data set 
and the distribution of the sensitive attribute within a
cluster of records 
is measured using Expression (\ref{eq:distance}),
then the data set satisfies differential privacy in the sense stated
in the following proposition.

\begin{proposition} \label{prop3} Let $k_{I}(D)$ be the function
that returns the view on subject $I$'s sensitive
attributes given a
data set $D$. If $D$ satisfies $\exp(\varepsilon/2)$-closeness
when using the distribution distance of Expression (\ref{eq:distance}),
then $k_{I}(D)$ satisfies $\varepsilon$-differential privacy.
In other words, if we restrict the domain of $k_{I}$ to $\exp(\varepsilon/2)$-close
data sets, then we have $\varepsilon$-differential privacy for $k_{I}$.
\end{proposition}

Proposition~\ref{prop3} is helpful to explain differential
privacy in terms of an intruder's knowledge gain 
on the sensitive attribute values of a target respondent
if the intruder can determine the target respondent's cluster.
This is shown in the next example.

\begin{example}
\label{ex2}
If one uses $\epsilon$-differential privacy with 
$\epsilon=2$, by Proposition~\ref{prop3} the probability weight
attached to a certain value of a sensitive attribute $X$ 
can grow by a factor of $e \approx 2.718$ if the
target individual's cluster is learned by the intruder. 
If the probability attached to a sensitive value in the cluster-level
distribution is deemed too high, then one needs to 
reduce $\epsilon$.
\end{example}

To decide whether a probability has grown too much,
one can resort to the connection with deniability highlighted
in Section~\ref{deny}, as follows:
\begin{itemize}
\item Consider that the reported value $v$ is now the 
cluster identifier.
\item Consider that the probabilities 
$\hat{p}_{vu} =\Pr(X=u|Y=v)$, for 
$u=1,\ldots,r$ are the probabilities assigned by 
the cluster-level distribution to the values of the sensitive
attribute within the cluster (we assume without 
loss of generality that the cluster contains $r$ different values).
\end{itemize}
With the above considerations, the problem of determining
the real value $X$ given the reported value $Y$ becomes 
the problem of finding the target respondent's sensitive value $X$
given the target respondent's cluster $Y$. 
Thus, we can use the following deniability argument 
to assess whether 
the cluster-level distribution has become too inhomogeneous:

\begin{example}
Assume the sensitive attribute can take $r=5$ different values
and that its data set-level empirical distribution 
is uniform, so that the relative frequency of each 
value is $1/5$. 
Take $\epsilon=2$ as in Example~\ref{ex2}. 
A cluster-level distribution where 
one value has relative frequency
$1/5 \times \exp(1) = 0.5436$ and the remaining four
values have relative frequencies $0.1141$ satisfies
$\exp(1)$-closeness; therefore, according 
to Proposition~\ref{prop3}, it satisfies 
$2$-differential privacy. However, whereas an intruder
cannot guess the sensitive attribute value
for a target respondent upon seeing the data set-level
distribution, the guess is much easier if the intruder
knows the target respondent's cluster, because one 
of the values concentrates more than 50\% of the relative
frequency. Thus, $\epsilon=2$ does not seem to 
offer enough protection.
\end{example}  

\section{PRAM and the permutation paradigm}
\label{pramperm}

In~\cite{DomiMura}, a permutation paradigm of anonymization was 
introduced. The authors first presented the following 
algorithm, that considers in turn each attribute $X$ 
in an original data set ${\bf X}$ and the corresponding attribute $Y$
in the anonymized data set ${\bf Y}$, and outputs an attribute $Z$ 
that is called the {\em reverse-mapped} version of $X$:

\begin{algorithmic}[l]
\REQUIRE Original attribute $X=\{x_1, x_2, \cdots, x_n\}$
\REQUIRE Anonymized attribute $Y=\{y_1, y_2, \cdots, y_n\}$
\FOR{$i=1$ to $n$}
\STATE Compute $j=\mbox{Rank}(y_i)$
\STATE Set $z_i = x_{(j)}$ (where $x_{(j)}$ is the value of $X$ of rank $j$)
\ENDFOR
\RETURN $Z=\{z_1,z_2, \cdots, z_n\}$
\end{algorithmic}

From the algorithm description, $Z$ is a permutation of $X$,
and the rank order of $Z$ is the same as the rank order of $Y$.
Since the above algorithm makes no assumption on the anonymization
procedure being used, it follows that 
any microdata anonymization technique
 is {\em functionally equivalent} to performing 
the following two steps one after the other:
\begin{enumerate}
\item {\bf Permutation}. Each attribute $X$ of the original
dataset is permuted into the corresponding $Z$.
Thus, the data set ${\bf X}$ is transformed into 
a data set ${\bf Z}$. 
\item {\bf Residual noise addition}. Noise is added
to each value of ${\bf Z}$ to obtain the anonymized data set
${\bf Y}$ (the noise is residual, because the ranks
of ${\bf Z}$ and ${\bf Y}$ must stay the same).
\end{enumerate}

Let us now look at how PRAM fits in the permutation paradigm.
PRAM does not permute in the sense of swapping attribute
values between records in the data set. Rather, it permutes
in the {\em domain} of attributes: that is, a data set 
anonymized with PRAM may contain attribute values not present
in the original data set. Hence, in terms of the permutation
paradigm, PRAM should be viewed as permutation plus some noise.
However, if ${\bf Y}$ is a PRAM-ed data set, it can be 
reverse-mapped to a permuted data set ${\bf Z}$ as explained
above. 

\section{Calibrating anonymization}
\label{calibrate}

In Section~\ref{rrpram} we connected PRAM and randomized
response, in Section~\ref{rrdp} we connected randomized response
and differential privacy, in Section~\ref{dpt} we connected
differential privacy and $t$-closeness and in Section~\ref{pramperm}
we connected PRAM and the permutation paradigm. 
Therefore, the privacy notions captured by all those 
privacy models are less different than it seems. In particular,
all of them can be expressed in terms of {\em any of the two 
following basic privacy ideas: deniability and permutation}.

As illustrated in the examples above, deniability is useful
to understand the privacy implications of relaxing the 
privacy parameters of $\epsilon$-differential privacy or $t$-closeness
in quest of utility.

Permutation may also be useful to understand the level of 
privacy achieved and calibrate the anonymization parameters
suitably. In~\cite{Muralidhar}, we showed how to use the permutation
distance to find anonymization parameters that make any  
linkage claimed by an intruder between the anonymized data set and an external
identified data set plausibly deniable by the data controller. In general,
the stronger the anonymization, the more similar
the permutation distances to those of random permutation, 
and the more linkage deniability. 
In the extreme case, a very strong anonymization can be 
viewed as a random permutation yielding random-looking
data; any linkage between random-looking data and original data 
is intrinsically deniable. 

We next explore two tools to calibrate permutation
to attain suitable protection against disclosure and
acceptable information loss. Due to the connections 
between permutation and the main privacy models shown in 
this paper, calibrating permutation can have broad applicability.

\subsection{$({\bf d}, {\bf v}, f)$-permuted privacy}

A privacy model called $({\bf d}, {\bf v}, f)$-permuted privacy
inspired in the above permutation paradigm was proposed
in~\cite{DomiMura}.
Given a vector ${\bf d}=(d^1,\ldots, d^m)$ of non-negative
integers, a vector ${\bf v}=(v^1,\ldots,v^m)$ of non-negative
real numbers, an original data set ${\bf X}$ and an anonymized
data set ${\bf Y}$ both with $m$ attributes, and a record-level
mapping $f:{\bf X} \longrightarrow {\bf Y}$,
we say ${\bf Y}$ satisfies
$({\bf d},{\bf v},f)$-permuted privacy {\em with
respect to original record ${\bf x}=(x^1, \ldots, x^m) \in {\bf X}$} if
$y^j_*$ being the value of the $j$-attribute $Y^j$ in the anonymized
data set closest to $x^j$ for $j=1,\ldots,m$,
\begin{enumerate}
\item The anonymized record $f({\bf x})=(y^1, \ldots, y^m)$ satisfies
\[ |\mbox{Rank}(y^j)-\mbox{Rank}(y^j_*)| \geq d^j \;\;\;(j=1,2,\ldots, m)\]
($d^j$ is called the {\em permutation distance} for the $j$-th attribute);
\item If $S^j(d_j)$ is the set of
values of the sorted $Y^j$ whose rank differs no more than $d_j$ from the rank
of $y^j_*$,
then the diversity of $S^j(d_j)$ is greater than $v^j$ according
to a given diversity criterion.
\end{enumerate}

If anonymization is just a permutation, then $y^j_*=x^j$.
For each original record ${\bf x}$, the data protector
can take as $f({\bf x})$ the anonymized record derived
from ${\bf x}$.
Diversity criteria for $S^j(d_j)$ may be the variance,
one of the $l$-diversity criteria, or the $t$-closeness criterion.
If $({\bf d}, {\bf v}, f)$-permuted privacy holds w.r.t. all
records in ${\bf X}$, then we say it holds for the dataset ${\bf X}$.

If an anonymization method $M$ with parameter $parms$ 
is used to obtain the anonymized
 version ${\bf Y}$ of an original data set ${\bf X}$, 
the data protector can compute what 
values of ${\bf d}$ and ${\bf v}$ 
the  $({\bf d}, {\bf v}, f)$-permuted privacy model holds with.
If the variabilities in ${\bf v}$ are not deemed sufficient
or if the permutation distances in ${\bf d}$ are not enough
to provide deniable linkages, then the anonymization parameters
$parms$ should be stronger or the method $M$ should be changed. 

\subsection{Aggregation measures based on power means}

Also drawing inspiration on the permutation paradigm, 
an aggregation measure based on power means was proposed 
in~\cite{NicolasInfSciences} 
to aggregate the absolute permutation distances $p_1, \ldots, p_n$ 
resulting from anonymizing the values of an attribute
in the $n$ records of a data set:
\begin{equation}
\label{powermean}
 J((p_1,\ldots, p_n),\alpha) = \left\{ \begin{array}{ll} 
\left(\frac{1}{n}\sum_{i=1}^n p^{\alpha}_i \right)^\frac{1}{\alpha} & 
\mbox{for $\alpha \neq 0$;}\\
\Pi_{i=1}^n p_i^{\frac{1}{n}} & \mbox{for $\alpha=0$,}
\end{array}\right. 
\end{equation}
where $\alpha < 1$ turns the above measure
into a disclosure risk measure and $\alpha > 1$ into an 
information loss measure.
Indeed, the more $\alpha$ approaches $-\infty$, the greater
is the weight of smaller permutation distances
in Expression (\ref{powermean}); since disclosure occurs 
when permutation distances for some values are too small, 
we have a disclosure risk measure when $\alpha$ is small.
On the other hand, the more $\alpha$ approaches 
$+\infty$, the greater is the weight of larger permutation
distances in Expression (\ref{powermean}); since large
permutation distances are the ones that most deteriorate
utility, we have an information loss measure when
$\alpha$ is large.
Thus, for $\alpha < 1$, the greater the value of  
of $J((p_1,\ldots, p_n),\alpha)$, the more disclosure risk,
whereas, for $\alpha > 1$, the greater the value 
of $J((p_1,\ldots, p_n),\alpha)$, the more information loss.  

From the above discussion, when $\alpha < 1$, we have that
$\alpha$ behaves as a risk aversion parameter: 
the smaller the $\alpha$ value chosen by the data controller, 
the more intolerable is disclosure considered. Analogously,
when $\alpha >1$, we have that $\alpha$ behaves as an information loss 
aversion parameter: the larger the $\alpha$ chosen by 
the data controller, the more intolerable is information loss.

Thus, if the controller is able to parameterize her risk aversion
by choosing $\alpha_1 < 1$ and her information loss aversion 
by choosing $\alpha_2 > 1$, she can use $J((p_1,\ldots, p_n),\alpha_1)$
and $J((p_1,\ldots, p_n),\alpha_2)$ to calibrate the anonymization
method $M$ and its parameters $parms$. 
Admittedly, choosing the right $\alpha_1$ and $\alpha_2$ 
and assessing whether $J((p_1,\ldots, p_n),\alpha_1)$ 
and $J((p_1,\ldots, p_n),\alpha_2)$ are acceptable may not 
be intuitive in most cases.

Anyway, these power-means measures may be used
to compare the disclosure protection and the information
loss achieved by two different anonymization methods $M$ and $M'$
(or by the same method $M$ with different parameters $parms$
and $parms'$). 

\section{Conclusions and further research}
\label{conclude}

We have highlighted several connections between privacy models,
which gives a deeper insight into their nature.
 They turn out to be more related 
than anticipated, around the principles of deniability 
and permutation. 
In particular, randomized response turns out to be very 
modern in spite of it having been introduced more than 50 years
ago: it is a local anonymization method and it allows
understanding the protection offered by 
$\epsilon$-differential privacy when $\epsilon$ is increased
to improve utility. A similar understanding of the effects
of a larger $\epsilon$ can be gained by looking at the connection
between $\epsilon$-differential privacy and $t$-closeness
under the light of deniability. 
Finally, since PRAM can be viewed as permutation 
and is connected with randomized response
and differential privacy, and the latter is connected
to $t$-closeness, the permutation principle can be used
to explain the guarantees offered by all those models.
Hence, calibrating permutation is a matter of high
interest in anonymization, and we have sketched two 
ways of approaching this issue.

Future research will involve giving more detailed guidelines
for calibrating privacy models and anonymization methods 
in view of optimizing the trade-off between disclosure risk
and information loss. 
Also, the knowledge gained on 
the common underlying principles of those models and methods should
be helpful to tackle the 
grand challenge of applying and/or adapting them to big data.

\section*{Acknowledgments and disclaimer}

The following funding sources are gratefully acknowledged:
European Commission (projects H2020-644024 ``CLARUS'' and
H2020-700540 ``CANVAS''), Government of Catalonia
(ICREA Acad\`emia Prize to J. Domingo-Ferrer) and
Spanish Government (projects TIN2014-57364-C2-1-R ``SmartGlacis''
and TIN2015-70054-REDC).
The views in this paper are the authors' own and do not necessarily
reflect the views of UNESCO or any of the funders.

	\bibliographystyle{IEEEtran}

\end{document}